\RequirePackage[bookmarksnumbered,unicode]{hyperref} 
\documentclass[sigconf]{cidr-2027}

\usepackage{graphicx}
\usepackage{xcolor}

\begin{document}

\title{From Data Querying to Data Investigations:\\Rethinking Natural Language Interfaces for Databases} 

\author{Fabian Wenz}
\orcid{0009-0008-8419-7205}
\affiliation{
\institution{TU Darmstadt and MIT}
  \city{Darmstadt}
  \country{Germany}
}

\author{Zixuan Chen}
\affiliation{
\institution{TU Darmstadt}
  \city{Darmstadt}
  \country{Germany}
}

\author{Carsten Binnig}
\affiliation{
\institution{TU Darmstadt, hessian.AI and DFKI}
  \city{Darmstadt}
  \country{Germany}
}

\renewcommand{\shortauthors}{Wenz et al.}
\newcommand{\system}{\texttt{\(D^2\)}}
\newcommand{\fabian}[1]{\textcolor{blue}{#1}}

\begin{abstract}
Natural language (NL) interfaces to databases have been optimized for the wrong problem. The dominant Text-to-SQL paradigm assumes that users ask questions that can be answered by single SQL queries. In practice, however, users seek assistance with solving data problems. This requires searching a database by sequences of SQL queries while reasoning over intermediate results instead of just running one SQL query. 
This paper therefore introduces a new paradigm for NL interfaces to data, which we call \emph{data investigations}. 
We present \system{}, a first prototype of a data investigation system that embodies this vision by autonomously searching, reasoning over, and collecting data to solve data problems.
Using a newly constructed benchmark based on the \emph{Murder Mystery} dataset, we demonstrate the potential of \system{} for tasks that require data investigations with evidence-backed decisions, extending beyond the capabilities of traditional single-query question answering.
\end{abstract}

\maketitle

\section{Introduction}\label{sec:introduction}

\noindent\textbf{Text-to-SQL is the wrong abstraction.}
Natural language (NL) interfaces to databases have advanced rapidly in recent years. Their dominant formulation is \emph{Text-to-SQL}: a user asks a question in NL, and the system translates it into a SQL query whose result answers that question. Benchmarks such as Spider~\cite{yu2018spider} and BIRD~\cite{wang2023bird} have driven impressive progress under this paradigm. Yet they are built on a fundamental assumption. We argue that it is way too limiting to assume that users formulate requests that correspond to a single SQL query. In practice, they do not \cite{DBLP:conf/hilda/DorschnerJB26}. Users neither think in terms of SQL queries nor care whether their request requires one SQL statement or multiple ones. They simply expect a system to help them solve a problem using their data. 

\begin{figure}
    \centering
    \includegraphics[page=1,width=0.97\linewidth, trim={0 0 0 2cm},clip]{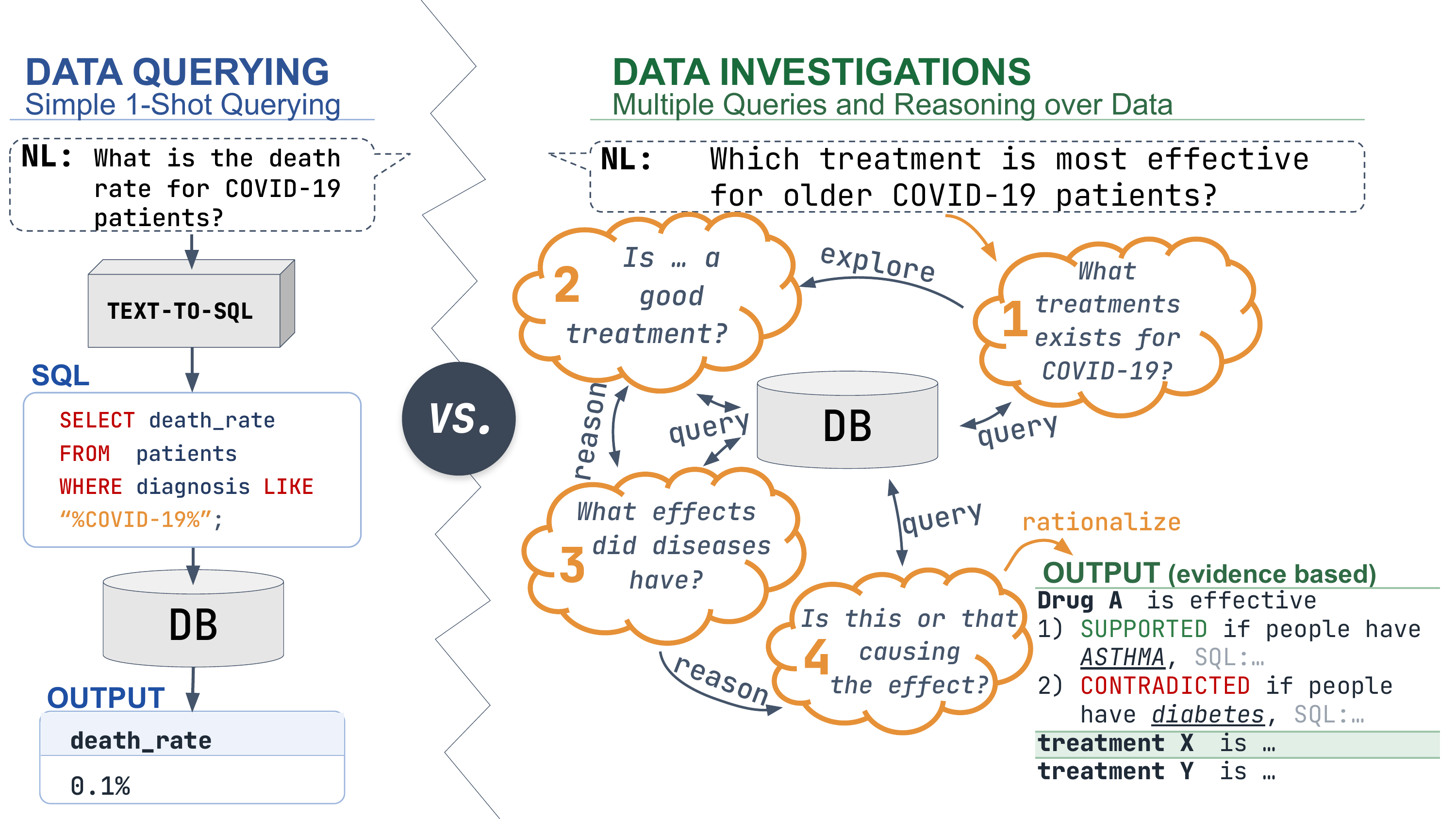}
    \vspace{-3ex}
    \caption{Comparison of traditional Text-to-SQL and Data Investigations. Text-to-SQL maps an NL question to a single SQL query, producing a direct result. In contrast, Data Investigations decompose a high-level NL question into multiple exploratory queries and reasoning steps, iteratively collecting and synthesizing evidence to verify or falsify potential solutions as output for the user question.}
    \label{fig:overview}
    \vspace{-6ex}
\end{figure}

\noindent\textbf{From querying data to investigating data.}
Consider a physician interacting with a hospital database. Asking for the average age of patients with COVID-19 is a valid database query, but it is rarely the question that matters. A physician is much more likely to ask: \emph{Which treatment is more effective for older COVID-19 patients?} Answering this question cannot be reduced to generating a single SQL query. Instead, the system must identify alternative treatment strategies, compare medications, account for differences in patient populations, analyze recovery outcomes, and reason about all factors contributing to successful treatments. Also, the outcome is not a single query answer but a collection of competing hypotheses (i.e., different treatments), each supported by different pieces of evidence. Recent studies on natural language interfaces suggest that this investigative interaction is what users actually expect: they seek assistance in solving data-driven problems, which we call \emph{data investigations}, rather than asking simple database queries~\cite{guo2023talk2data}.  
Figure~\ref{fig:overview} contrasts this data investigative paradigm with the traditional Text-to-SQL workflow.
By exploring alternative treatments, gathering supporting and contradicting evidence, and making the reasoning process transparent, data investigations enable users to understand, verify, and ultimately trust the system's conclusions. 

\begin{figure*}
    \centering
    \includegraphics[page=2,width=0.95\textwidth, trim={0 5.4cm 0 3.5cm},clip]{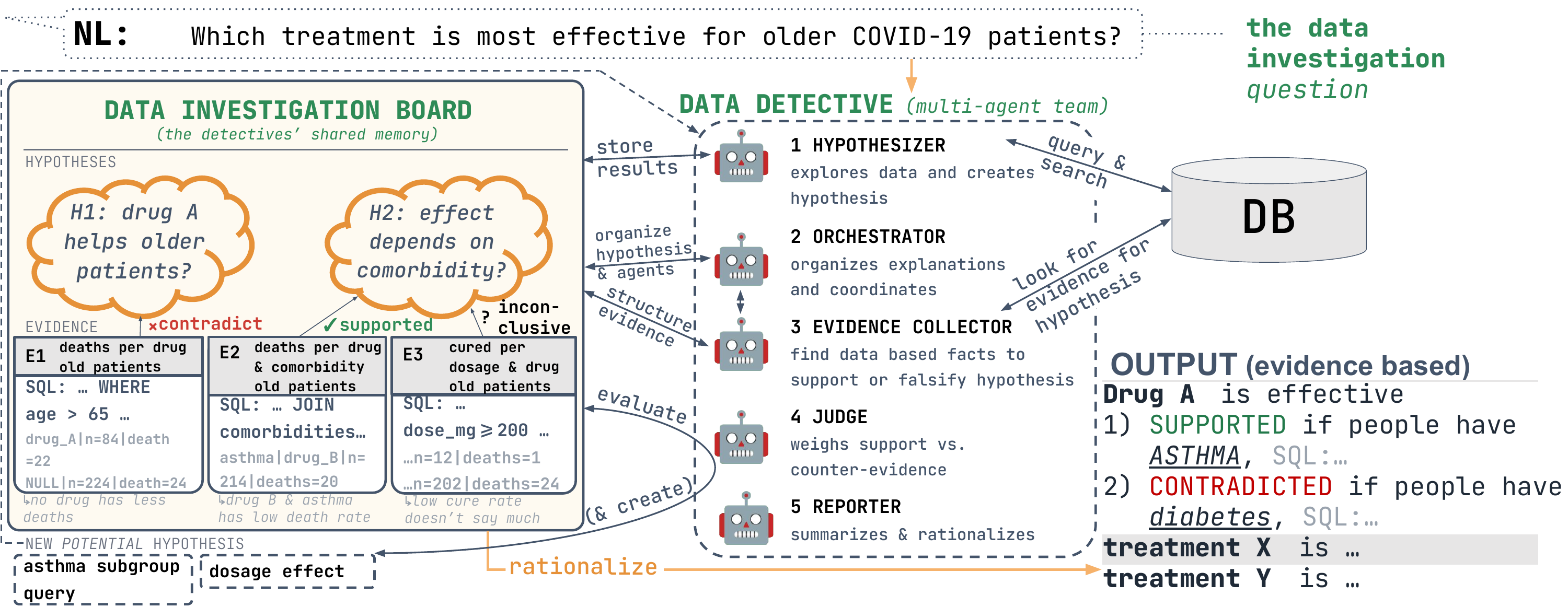}
    \vspace{-3ex}
    \caption{Architecture of a Data Investigation System with two main system abstractions: (1) The \emph{Data Detective}, a multi-agent team that iteratively inquires the database, formulates and evaluates hypotheses. (2) The team stores supporting and contradicting evidence on a shared \emph{Data Investigation Board} before producing an evidence-backed result. The result is the verdict on each hypothesis (supported or contradicted) and the summary of its evidence.}
    \label{fig:framework}
    \vspace{-4ex}
\end{figure*}

\noindent\textbf{This is a new data management workload.}
And the principle of data investigations extends far beyond healthcare. A business analyst may ask \emph{"why customer retention suddenly declined"}, a manufacturer may ask \emph{"where to build the next production facility"}, or a city planner may ask \emph{"why a traffic light behaved unexpectedly"}. None of these questions has a unique SQL query that provides the answer. Instead, the system must iteratively explore the available data, formulate possible hypotheses based on the data, collect supporting and contradicting evidence, revise its understanding as new observations emerge, and ultimately communicate well-supported conclusions to the user. We argue that this investigative process constitutes a fundamentally new data management workload that is distinct from translating natural language into SQL.

\noindent\textbf{Current systems are not designed for this workload.}
Today's Text-to-SQL approaches primarily optimize the translation from NL into SQL~\cite{yu2018spider,wang2023bird, chen2024beaver}. While this is an important capability, it addresses only one component of the broader problem of a data investigation. Investigative workloads require managing an evolving search over the data, involving sequences of queries whose execution depends on reasoning over intermediate results rather than generating a single query upfront.
One might argue that recent LLM agents equipped with database tools already provide this capability~\cite{DBLP:conf/cidr/EckmannB26,palimpzest,wenz2026rubicon}. 
Such systems represent an important step toward data investigations. However, as we show later in this paper, today's pure agentic approaches remain insufficient. First, agents typically tend to terminate early once they identify one plausible answer to a question, whereas many real-world data problems have multiple equally relevant solutions—for example, several effective treatments for COVID-19. Second, their reasoning process is often difficult for users to inspect as it is "buried" in a long text trace.

\noindent\textbf{Towards Data Investigation Systems.}
The central argument of this paper is that NL interfaces should no longer be viewed primarily as Text-to-SQL. The ultimate goal should instead be to help users solve problems based on the available data. 
We argue that this requires a new systems layer between users and relational databases, which we coin \emph{data investigation systems}. 
As a first step toward this vision, we present \system{}, a prototype that introduces two new system abstractions for hypothesis-driven data analysis: the \emph{Data Detective} and the \emph{Data Investigation Board}. The \emph{Data Detective} is a team of specialized agents that collaboratively investigates a data problem by generating hypotheses, and collecting supporting or contradicting evidence for each hypothesis --- i.e., how efficient different treatments are in our example. The \emph{Data Investigation Board} serves as the detective's workspace, externalizing the reasoning process as shared investigation state between agents. By making the investigation state a first-class systems object, \system{} enables systematic exploration of all alternative hypotheses and enables an easier verifiability of the outcome of an analysis for humans.

\noindent\textbf{Our contributions.} To summarize, as the main contribution, this paper introduces data investigations, a new paradigm for NL interfaces to data that shifts the focus from directly answering queries to solving more open-ended data problems. We present \system{}, a very first prototype realizing this paradigm. We further introduce a benchmark of data investigation workloads and use it to study the capabilities and limitations of current systems. Finally, we outline an agenda for building future data investigation systems.

\section{Our Vision: Data Investigation Systems}
\label{sec:vision}

Figure~\ref{fig:framework} illustrates our vision for future \emph{data investigation systems} using a running example where a clinician asks \emph{Which treatment is most effective for older COVID-19 patients?} A data investigation system acts as a \emph{detective for data}: it explores the database, develops hypotheses such as \emph{"Drug A helps older patients"} by looking at the data, and gathers evidence from data to support or falsify hypotheses. At the core, we argue that data investigation systems consist of two components: the \emph{Data Detective}, a multi-agent system with specialized roles for hypothesis generation, evidence collection, and hypothesis evaluation (i.e., the hypothesis is true or false based on the evidence) and the \emph{Data Investigation Board}, an externalized shared memory that organizes all state for data investigations, as shown on the left of the figure. 
In the following, we first discuss how data investigation workloads are different from existing workloads and then describe the design of the two main components to support such workloads.

\noindent\textbf{What are Data Investigations?}
A \emph{data investigation} differs fundamentally from Text-to-SQL. Rather than translating a NL question into a single SQL query, the system receives a high-level investigative objective, such as \emph{Which treatment is most effective for older COVID-19 patients?} in Figure~\ref{fig:framework}. At the start, neither the relevant queries nor the relevant tables may be known to answer such a user question. A data investigation system must therefore explore the data first and see what treatments are available in the data, which involves exploring the schema and executing several queries already. Based on this, a system can then formulate candidate hypotheses (e.g., H1 in Figure~\ref{fig:framework}).
However, finding "Drug A" as a treatment for older patients does not mean it is an effective treatment. As such, the system next needs to find evidence in the data for (or against) this hypothesis by further querying the data. 
In the example, the evidence E1, which is the result of a SQL query, contradicts H1 because treated and untreated patients over 65 show similar death rates.
Also H1 was only one hypothesis; as such, the system continues collecting evidence for the other hypothesis.
An interesting characteristic of a data investigation workload is that it progressively discovers relevant hypotheses and evidence through exploration, querying the data as well as reasoning to decide which hypotheses hold and which do not. 

\noindent\textbf{The Data Detective.}
The core component of our architecture of a data investigation system to conduct such a data investigation process is the \emph{Data Detective}. Given the high-level user query, it coordinates hypothesis generation, evidence collection, and evidence-based reasoning toward a final conclusion, such as that \emph{"Treatment A helps patients with asthma but not those with diabetes"}.
A simple implementation of a \emph{Data Detective} would be to equip an agent with a SQL tool and let it iteratively explore and query the database. However, this not only keeps the investigation process implicit inside the LLM context, making it hard for the agent to keep track of an investigation state, but it also causes the agent to prematurely converge on a single hypothesis rather than systematically explore all alternatives available in the data, as we show in our initial evaluation in Section \ref{sec:eval}. To overcome these limitations, we thus design our \emph{Data Detective} as a multi-agent system, which divides responsibilities of data investigation to avoid these shortcomings. As shown in Figure~\ref{fig:framework}, each agent has a specialized role: a \emph{Hypothesizer} independently generates candidate hypotheses (H1, H2) from data and thus guarantees a breadth of evaluation, while an \emph{Orchestrator} coordinates the investigation and oversees that all hypotheses will be checked based on the available data. Furthermore, a team of \emph{Evidence Collectors} gathers supporting or contradicting data as evidence for or against a hypothesis, while a \emph{Judge} evaluates the evidence and comes to a verdict per hypothesis. Finally, once all hypotheses are checked, a \emph{Reporter} summarizes the collected evidence for users to understand the reasoning process. 

\noindent\textbf{The Data Investigation Board.}
The second key component is the \emph{Data Investigation Board}, the externalized memory of the agents shown on the left of Figure~\ref{fig:framework}. Instead of hiding the investigation state in the context of an LLM, the board explicitly represents the state of evolving investigation, linking hypotheses to evidence that was found so far to support or contradict hypotheses. 
While recent work on agentic context management advocates externalizing memory to efficiently support long-horizon reasoning~\cite{li2026acmagenticcontextmanagement}, the \emph{Data Investigation Board} goes beyond this and structures the memory in particular for data investigations.
Important is that this structure is problem-independent and can be used for data investigations in any domain.
Moreover, the investigation board provides a persistent representation that can be shared across all agents to coordinate investigations and also resume in case agents fail (which happens). In addition, it also supports parallel investigative branches (e.g., different collectors working on different hypotheses), but it also preserves all outcomes of hypotheses (rejected and supported ones), and allows users to understand not only the final conclusion but also the evidence and reasoning that led to it.

\begin{figure*}
\centering
\includegraphics[page=3,width=0.97\linewidth, trim={0 5cm 0 0.9cm},clip]{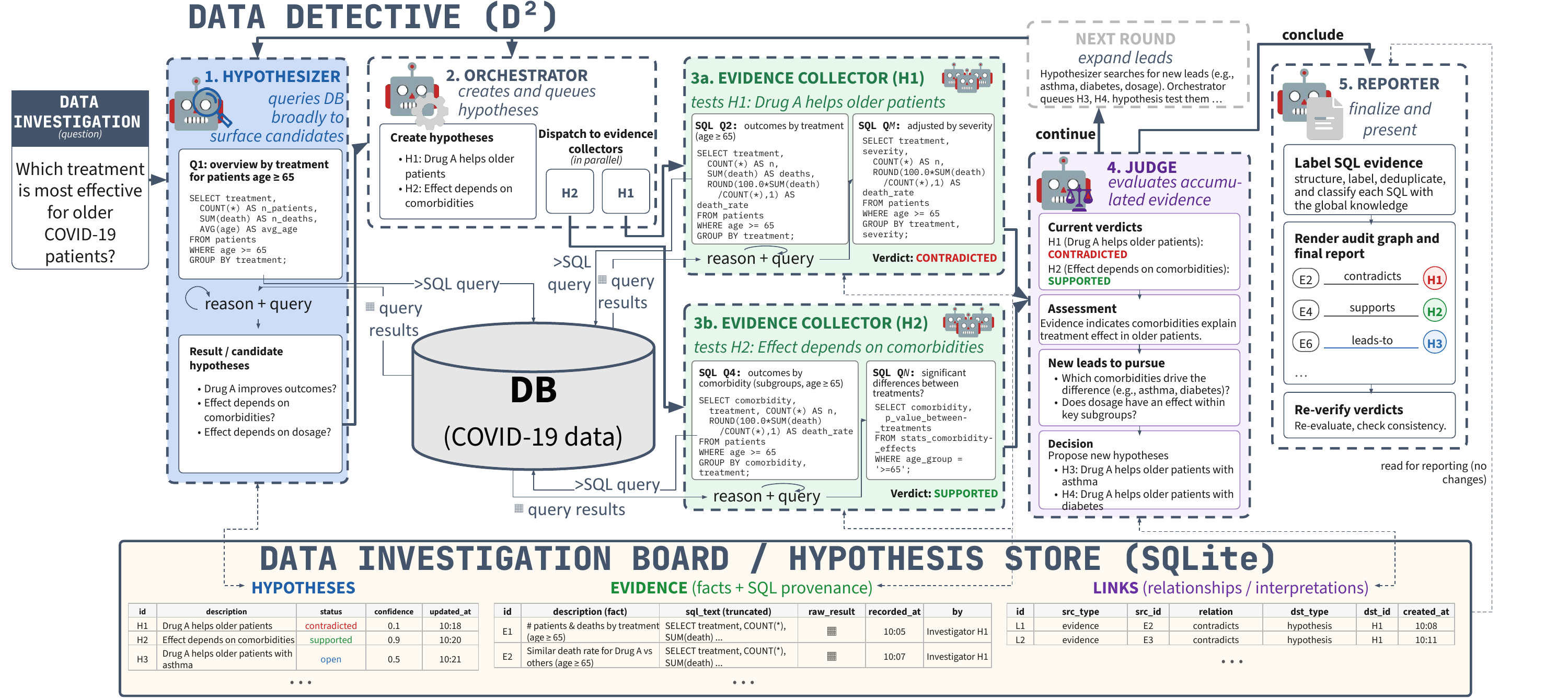}
\vspace{-3ex}
\caption{Our \system{} prototype in action. An investigation over a COVID-19 patient database answering ``Which treatment is most effective for older patients?''. The \emph{Data Detective} (top, steps 1--5) is a multi-agent system: the hypothesizer (1) explores the database and proposes hypotheses (e.g., Drug~A benefits older patients; effects depend on comorbidities); the orchestrator (2) assigns them to evidence collectors (3a, 3b), which query the database in parallel for evidence and counter-evidence; the judge (4) evaluates evidence and comes to a verdict, ($\rightarrow$1/2) it also refines hypotheses if needed, or triggers (5) the reporter to produce a provenance-backed report. The \emph{Data Investigation Board} (bottom) provides the shared state, enabling consistent collaboration across agents and investigation rounds independently for all hypotheses.}
\vspace{-4ex}
\label{fig:agent-graph}
\end{figure*}

\vspace{-3ex}
\section{Our Prototype: The \system{} System}
\label{sec:prototype}

Based on the architecture outlined before, we have been building a first data investigation system we call \system{}. 
Figure~\ref{fig:agent-graph} shows a representative run of \system{} over a COVID-19 patient database.
We want to highlight that the workflow of \system{} is completely domain-independent. Replacing the COVID-19 database with a police, manufacturing, or any other database leaves the investigation process completely unchanged.
In the following, we discuss in detail how \system{} works as well as the technical challenges we faced when building it.

\vspace{-2ex}
\subsection{The Data Detective}
\label{sec:detective}

Unlike a ReAct-style agent that incrementally generates SQL until it reaches a first plausible answer, the task of the \emph{Data Detective} is to systematically analyze and collect data-based evidence for all hypotheses that can be derived from the data. 
As shown in Figure~\ref{fig:agent-graph}, the \emph{Data Detective} decomposes the investigation into five specialized agent roles.
It is important that all agents work fully independently with the data, and the steps outlined below are not a strict execution sequence.
Instead, agents collaborate and share state through the shared \emph{Data Investigation Board} with clear interfaces for how they can contribute to or read from the board.

\noindent\textbf{Hypothesizer (step~1).}
Every investigation begins with understanding what potential answers to the user query might be.
The role of the hypothesizer, however, is not to produce a single answer, but to explore the search space of possible hypotheses. 
For this, the hypothesizer agent has an SQL tool to explore the schema and execute queries.
Given an investigative objective, it uses these tools and, based on the data, it proposes multiple candidate hypotheses that deserve further investigation by adding them to the investigation board. In our example, the hypothesizer first issues a broad query \texttt{Q1} that computes statistics for all treatments for patients aged 65 and older, and from the result proposes three competing candidates: Drug~A improves outcomes, the effect depends on comorbidities, and the effect depends on dosage. 

\noindent\textbf{Orchestrator (step~2).}
The orchestrator provides the global control loop of the investigation. 
It picks hypotheses and decides which hypotheses should be investigated next, which should be abandoned (as it is not really answering the user query), and when sufficient evidence has accumulated to terminate the investigation. In the example, it selects two of the hypothesizer's candidates -- H1 (``Drug~A helps older patients'') and H2 (``effect depends on comorbidities'') -- and dispatches them to independent evidence collectors in parallel. Importantly, the orchestrator never plans and executes any SQL queries on the data. Instead, it reasons entirely at the level of all hypotheses which are available in the investigation board.

\noindent\textbf{Evidence collectors (steps~3a+b).}
Each hypothesis is assigned its own evidence collector. Collectors independently explore the database by issuing SQL queries to collect data-based evidence that helps to support or reject a hypothesis and adds the evidence to the \emph{Data Investigation Board}. In the example, the collector assigned to H1 issues \texttt{Q2} and \texttt{Q3} to compare death rates by treatment and by severity, and concludes that H1 is \emph{contradicted}; concurrently, the collector assigned to H2 issues \texttt{Q4} and \texttt{Q5} to compare outcomes across comorbidity subgroups. Since collectors share persistent investigation state via the board rather than holding it in their private transient reasoning context, multiple hypotheses can be investigated concurrently without interfering with one another.

\noindent\textbf{Judge (step~4).}
Once a collector decides sufficient evidence has been collected, it calls the judge. Continuing the example, the judge in our case weighs the two verdicts based on the collected evidence. This does not involve any database queries, but the main task of the judge is to reason over the evidence on the board and decide if a hypothesis is confirmed or contradicted based on the evidence. 
However, a judge can also decide that additional evidence needs to be collected by starting the evidence collection again with hints of what is missing, just like in a real investigation.
Moreover, a judge can also identify new potential hypothesis that stems from the verdict and the collected evidence. For example, in our case, as efficiency seems to depend on comorbidities, the judge proposes two refined hypotheses, H3 (``Drug~A helps older patients with asthma'') and H4 (``Drug~A helps older patients with diabetes'') which are processed by the agents as described before. 

\noindent\textbf{Next round ($\rightarrow$ step 1/2).}
When the judge creates new potential hypotheses instead of concluding, the orchestrator queues them  -- here H3 and H4 -- for a further round of evidence collection, closing the loop back to step~1 and~2. This is what allows the investigation in Figure~\ref{fig:agent-graph} to refine ``effect depends on comorbidities'' into concrete, testable subgroup hypotheses.

\noindent\textbf{Reporter (step~5).}
Finally, once the investigation for all hypotheses terminates, the reporter summarizes all investigations from the \emph{Data Investigation Board} with the goal of making it easy for the user to verify the reasoning process based on the hypothesis and evidence.
More precisely, the reporter therefore cleans redundant evidence, renders an audit graph connecting hypotheses to their cleaned evidence (e.g., evidence \texttt{E2} \emph{contradicts} H1, while \texttt{E4} \emph{supports} H2).
The goal of this audit graph is to be easily comprehensible by humans. 
As such, it is designed to be concise and structured.
This is in stark contrast to a reasoning trace that agents use, which is purely text-based and really hard for humans to understand easily.

\vspace{-1.5ex}
\subsection{The Data Investigation Board}

The second core component of a data investigation is the \emph{Data Investigation Board} which we have also implemented as part of \system{}, shown at the bottom of Figure~\ref{fig:agent-graph}. The board, as discussed before, maintains the externalized state of the investigation and also acts as the interface between agents that read from and write to the board. In our prototype, the board is implemented as a SQLite database with three tables centered around the main concepts of a data investigation. 

\noindent\textbf{The three investigation tables.}
The \emph{Hypotheses Table} stores candidate hypotheses together with their status and confidence, allowing \system{} to explore competing hypotheses without prematurely converging to one. The \emph{Evidence Table} records each observation along with its SQL query, a textual description of the query, the query result, and provenance (timestamp and investigator), making findings traceable. The \emph{Links Table} connects hypotheses and evidence, and a state capturing whether evidence supports or contradicts a hypothesis. Together, these tables store all relevant state of investigations and thus separate state from execution in the detective. Moreover, the board also preserves the complete investigation history.

\noindent\textbf{How does the board help?}
By externalizing hypotheses, evidence, and their relationships, it provides a shared representation that enables agents to collaborate across investigation rounds while each agent can focus on its particular role. 
We show in our initial evaluation that state-of-the-art agents that keep everything in their context lead to much lower coverage of which hypotheses are being explored, and also the reasoning process is hard to follow by humans. 
Using the board, users can audit the investigation directly from the board without inspecting the underlying LLM reasoning traces, which are hard to comprehend cognitively by humans, as all reasoning is hidden in a long unstructured text trace.
Moreover, the board makes long-running investigations resumable if agents crash or facts change.

\section{Initial Evaluation}
\label{sec:eval}

\noindent\textbf{Benchmark.}
As no benchmark exists for data investigations, we introduce a new benchmark inspired by the SQL Murder Mystery\footnote{\url{https://mystery.knightlab.com/}} where the goal is to find a murderer given the facts in a database. However, unlike the original benchmark, which contains a single murderer case, our benchmark comprises 50 fictional murder cases with varying complexity. 
The task is to identify the murderer(s) per case and produce an evidence-backed hypothesis.
The benchmark captures the core characteristics of data investigations: agents must first explore the database to discover relevant information, develop and refine competing hypotheses (i.e., who are potential murderers), and collect evidence (or counter-evidence) which is distributed across multiple tables in the entire database. Thus, success requires more than generating one correct SQL query; it requires a multi-step investigation that combines data exploration, reasoning, and evidence validation. 
Furthermore, we added cases of different difficulty which is determined by the combined number of difficulty dimensions (e.g., number of hints per suspect, normalized number of suspects per case, number of missing clues, and number of unreliable witnesses), yielding overall four categories: \emph{easy}, \emph{medium}, \emph{hard}, and \emph{very hard}. Harder cases require investigating more suspects, following longer chains of evidence, and, in some cases, identifying multiple suspects as murderers in one case. For checking quality, each case includes the ground-truth murderer set and the evidence required to solve the case (which we manually annotated), enabling evaluation of both the final verdict and the quality, completeness, and provenance of the investigation.

\noindent\textbf{Baselines.}
We compare \system{} against Claude Code and OpenClaw, two general-purpose agentic systems supporting iterative SQL tool use to explore and query the database. All three systems (\system{} and the baselines) use Claude Sonnet 5 as their underlying model and receive identical case descriptions, schemas, database access, and instructions, and may iteratively issue SQL queries before producing a final answer. All collected evidence is kept in the system context. 

\noindent\textbf{Metrics.}
We evaluate investigations along three dimensions:

\noindent\emph{Correctness} measures whether the system found the right answer (i.e., who was the murderer?). It is not about whether all potential suspects (i.e., hypotheses) were analyzed, which is the goal of completeness. Correctness is measured as F1 over precision (did we find the correct murderers?), which penalizes false accusations, and recall, which penalizes missed perpetrators (did we miss murderers?).

\noindent\emph {Completeness} measures whether the system explores the full set of all hypotheses (i.e., all suspects). A good data investigation system should explore them all, as sometimes multiple possible solutions exist, which is a big difference to querying data. 
We report the F1 score of hypotheses analyzed over the set of true hypotheses.

\noindent\emph{Verifiability Ease} measures how easily humans can understand and verify the reasoning of a data investigation system. Systems that explicitly structure their reasoning, for example by linking hypotheses to evidence, reduce the human effort required for verification. In contrast, approaches such as Claude Code or OpenClaw represent reasoning as unstructured text. Formally, the metric measures graph similarity between the generated reasoning structure and an ideal hypothesis-evidence-graph by solving a bipartite matching problem. Due to space constraints, we omit the formal definition and provide it in the technical report, which we will publish on arXiv.

\begin{figure}
    \centering
    \includegraphics[width=0.97\columnwidth, trim={0 0 0 0},clip]{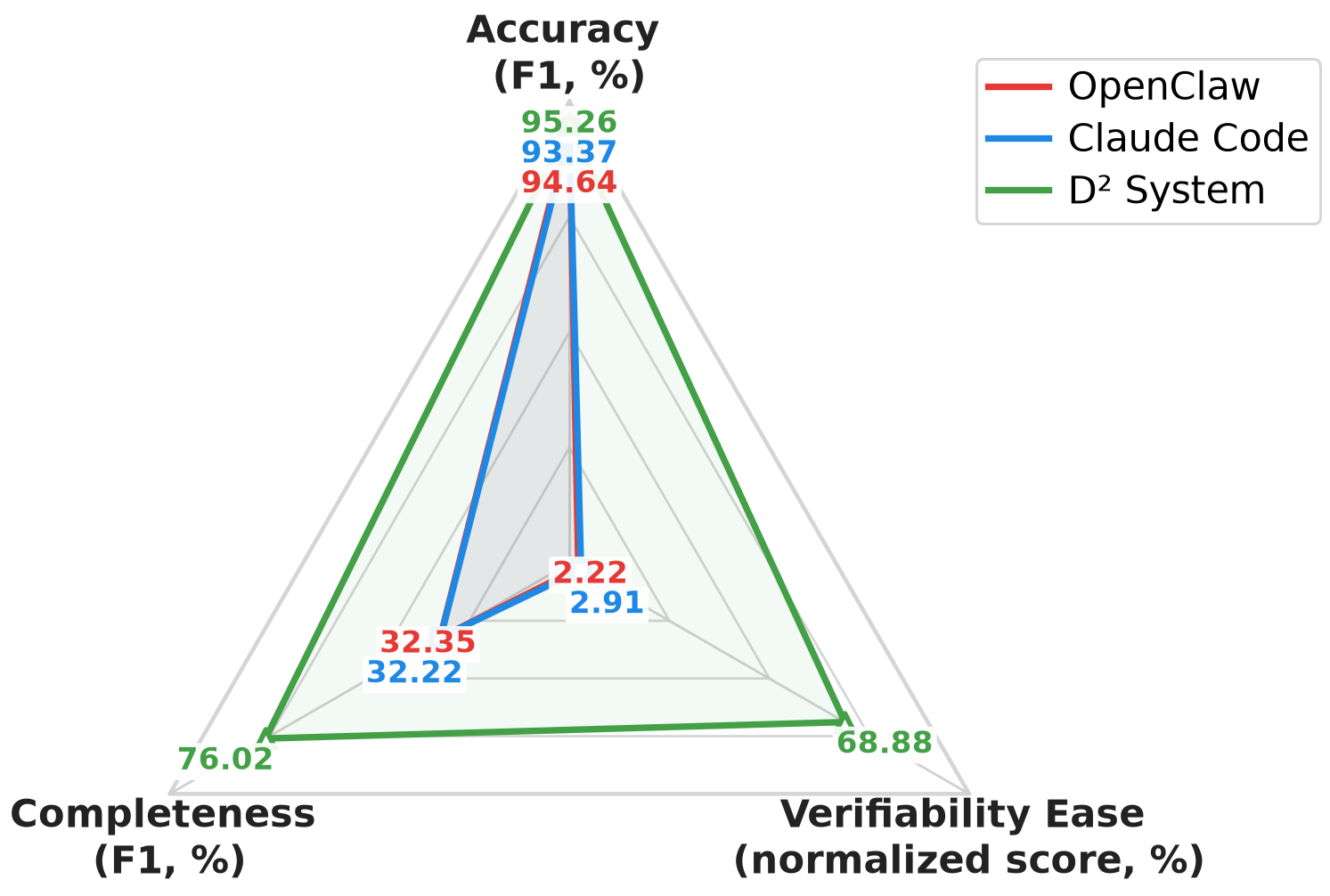}
    \vspace{-3ex}
    \caption{Overall investigation quality.
    Correctness, completeness, and verifiability ease. Higher-is-better. \system{} excels in all dimensions over baselines.}
    \vspace{-5ex}
    \label{fig:overall-results}
\end{figure}

\begin{table*}
    \small
    \centering
    \caption{Correctness, Completeness \& Verification Ease by case difficulty.}
    \vspace{-3ex}
    \label{tab:difficulty-results}
    \begin{tabular}{lccc|ccc|ccc|ccc}
        \toprule
        & \multicolumn{3}{c|}{Easy}
        & \multicolumn{3}{c|}{Medium}
        & \multicolumn{3}{c|}{Hard}
        & \multicolumn{3}{c}{Very Hard} \\
        System
        & Corr. 
        & Comp. 
        & Ver. Ease 
        & Corr. 
        & Comp. 
        & Ver. Ease 
        & Corr. 
        & Comp. 
        & Ver. Ease 
        & Corr. 
        & Comp. 
        & Ver. Ease  \\
        \midrule
        OpenClaw
        & \textbf{100.00\%} & 32.64\% & 2.30\%
        & 98.31\% & 30.66\% & 2.18\%
        & 93.55\% & 30.90\% & 2.15\%
        & \textbf{84.34\%} & 35.25\% & 2.27\% \\

        Claude Code
        & \textbf{100.00\%} & 42.44\% & 2.80\%
        & 98.31\% & 33.21\% & 2.84\%
        & 93.55\% & 25.31\% & 2.88\%
        & 79.47\% & 25.65\% & 3.12\% \\

        \system{}
        & \textbf{100.00\%} & \textbf{86.06\%} & \textbf{74.01\%}
        & \textbf{100.00\%} & \textbf{75.50\%} & \textbf{69.13\%}
        & \textbf{95.24\%} & \textbf{69.42\%} & \textbf{65.85\%}
        & \textbf{84.15\%} & \textbf{71.07\%} & \textbf{66.21\%} \\
        \bottomrule
    \end{tabular}
    \vspace{-3ex}
\end{table*}

\vspace{-1ex}
\subsection{Exp. 1 \& 2: Overall Quality \& Breakdown}

Our first experiment evaluates the end-to-end investigation quality of \system{}, Claude Code, and OpenClaw across all 50 cases in our benchmark.
For every system, we report the average correctness, completeness, and verifiability ease. Figure~\ref{fig:overall-results} shows the results - higher is better along all dimensions. All three systems identify the correct murderer: \system{} reaches 95.26\% correctness, compared with 94.64\% for OpenClaw and 93.37\% for Claude Code. If correctness was the only metric, one might conclude that all systems are equivalent.
They are not.
\system{} achieves 76.02\% completeness, while both baselines remain near 32\%. In other words, the baselines often find a plausible culprit and stop; \system{} instead is the only system that investigates the larger hypothesis space. The difference is even larger for verifiability ease: 68.88\% for \system{}, compared with 2.22\% for OpenClaw and 2.91\% for Claude Code. The baselines produce walls of text. \system{} produces a concise and structured audit graph.

Our second experiment examines how investigation quality changes as the difficulty of the investigation increases.
Table~\ref{tab:difficulty-results} shows the expected trend: correctness declines as cases become harder. All systems solve the easy cases, while performance drops to roughly 79--84\% on very hard cases. \system{} is therefore not immune to difficult investigations.
The more interesting result is what happens beyond correctness. Across every difficulty level, \system{} remains substantially more complete and more verifiable than the baselines. Even on the very hard cases, it reaches high completeness while OpenClaw and Claude Code remain low or even degrade. \system{}'s verifiability ease also remains above 65\%, while both baselines stay below 3\%.
Overall, \system{} is comparable in correctness but clearly outperforms the baselines in completeness and verification ease.

\vspace{-1ex}
\subsection{Exp. 3: Ablation Study}

Our final experiment studies which parts of our architecture contribute to the performance of \system{}.
For this, we collapse specialized agent roles and compare the resulting variants with the full \system{} system. This preserves the same tools, data, and general execution capabilities while changing how investigative responsibilities are distributed.
Moreover, all variants have access to the same investigation board.
We evaluate the following configurations: 

\noindent\emph{(1) Full \system{}:} the complete architecture with all roles.

\noindent\emph{(2) Without Reporter:} the final answer is generated directly from the accumulated evidence, without a reporter that cleans up evidence. We expect lower verification ease.

\noindent\emph{(3) Without Judge:} The verdict per hypothesis is taken by the evidence collector instead of a separate judge, which critically looks at the evidence. Also no new refined hypotheses are created. We expect lower completeness and correctness.

\noindent\emph{(4) Merged Hypothesizer and Orchestrator:} one agent both generates candidate hypotheses and manages the global investigation. We expect lower completeness.
    
\noindent\emph{(5) Single Agent (as Detective):} one agent manages all tasks, but we keep the Data Investigation Board. This should affect verification ease positively against baselines (Claude and OpenClaw) of Exp. 1.

\begin{table}[t]
    \small
    \caption{Ablation study of the \system{} system.}
    \vspace{-3ex}
    \label{tab:ablation-results}
    \centering
    \begin{tabular}{lccc}
        \toprule
        Configuration
        & Corr. 
        & Compl. 
        & Ver. Ease  \\
        \midrule
        \textbf{Full} \system{}
        & \textbf{95.26\%} & \textbf{76.02\%} & \textbf{68.88\%} \\

        Without Reporter
        & 89.58\%  & 73.54\% & 50.15\% \\

        Without Judge
        & 77.25\% & 61.19\% & 62.98\% \\

        Merged Hyp. \& Orch.
        & 89.67\% & 29.81\% & 39.21\% \\

        Single Agent
        & 86.63\% & 22.59\% & 21.34\% \\
        \bottomrule
    \end{tabular}
    \vspace{-6ex}
\end{table}

Table~\ref{tab:ablation-results} shows the results. Without the judge, irrelevant hypotheses survive, and new relevant ones are not created, reducing correctness and completeness. Verifiability is not that much affected due to the separate reporter. Without the reporter, the investigation remains largely intact, but its structure is lost, causing the strongest drop in verifiability ease. Merging the hypothesizer and orchestrator produces too few hypotheses, primarily reducing completeness, while verifiability and correctness remain comparatively stable. Finally, the single-agent variant approaches the baselines: completeness drops substantially. However, as we keep the board, verification ease is higher as for the baselines in Exp. 1. 

\vspace{-1.5ex}
\section{Research Roadmap}
This paper argues that data investigations constitute a new data management workload and presents \system{} as a first step toward supporting it. Building general-purpose data investigation systems raises fundamental questions about how investigations should be represented, planned, optimized, executed, and verified. We outline several additional promising directions below.

\noindent\textbf{Optimization of execution.}
Classical query optimization assumes a fixed declarative query and optimizes its execution plan~\cite{Chaudhuri98}. Data investigations introduce a broader optimization problem: the system must decide both \emph{what} to investigate and \emph{how} to execute it. Objectives extend beyond runtime to accuracy, completeness, verifiability, monetary cost, and resource consumption. An important research direction is therefore multi-objective investigation optimization, including cost models, adaptive investigation plans, and principled trade-offs between competing objectives.

\noindent\textbf{Provenance \& auditing.}
Investigation results require provenance connecting conclusions to the evidence and queries that produced them. The \emph{Data Investigation Board} provides a natural substrate for such provenance, extending classical ``why'' and ``why-not'' analysis to investigations. More broadly, \emph{reverse investigation} could use the investigation graph to identify missing, contradictory, or weak evidence on how an investigation reached an incorrect conclusion.

\noindent\textbf{Benchmarking.}
A key challenge is scaling from our prototype to realistic workloads and establishing more benchmarks for data investigation systems. This requires larger schemas, larger tables, more complex relationships, longer evidence chains, and more diverse questions, including open-ended ``why'' and ``how'' questions. Benchmarks with progressively more complex investigations are essential for evaluating systems beyond simple fact-finding tasks. As the number of agents grows, coordination also becomes more and more a systems problem.

\noindent\textbf{Heterogeneous data \& specialized operators.}
Our prototype currently relies primarily on SQL and tables as its investigation substrate. A general investigation system should integrate heterogeneous data and specialized operators such as images and text as well as optimization solvers and satisfiability engines as operators. This raises database-style questions around cost-based tool selection, operator composition, intermediate-result reuse, and provenance across heterogeneous execution engines.

\vspace{-2ex}
\bibliographystyle{abbrvnat}
\bibliography{bibliography-short.bib}

@misc{li2026acmagenticcontextmanagement,
      title={ACM: Agentic Context Management for Long Horizon Tasks}, 
      author={Xiaochuan Li and Ryan Ming and Meng Chu and Shuai Shao and Rong Jin and Chenyan Xiong},
      year={2026},
      eprint={2607.23809},
      archivePrefix={arXiv},
      primaryClass={cs.AI},
      url={https://arxiv.org/abs/2607.23809}, 
}

@misc{guo2023talk2data,
      title={Talk2Data: A Natural Language Interface for Exploratory Visual Analysis via Question Decomposition}, 
      author={Yi Guo and Danqing Shi and Mingjuan Guo and Yanqiu Wu and Qing Chen and Nan Cao},
      year={2023},
      eprint={2107.14420},
      archivePrefix={arXiv},
      primaryClass={cs.HC},
      url={https://arxiv.org/abs/2107.14420}, 
}

@inproceedings{DBLP:conf/hilda/DorschnerJB26,
  author       = {Charlotte D{\"{o}}rschner and
                  Frank J{\"{a}}kel and
                  Carsten Binnig},
  title        = {{Towards More Realistic Natural Language Queries in Text-to-SQL Benchmarks}},
  booktitle    = {Proceedings of the Workshop on Human-In-the-Loop Data Analytics, {HILDA}
                  2026},
  pages        = {1--8},
  publisher    = {{ACM}},
  year         = {2026}
}

@inproceedings{Chaudhuri98,
  author       = {Surajit Chaudhuri},
  title        = {An Overview of Query Optimization in Relational Systems},
  booktitle    = {Proceedings of the Seventeenth {ACM} {SIGACT-SIGMOD-SIGART} Symposium
                  on Principles of Database Systems},
  pages        = {34--43},
  publisher    = {{ACM} Press},
  year         = {1998}
}

@misc{wenz2026rubicon,
      title={RUBICON: Agentic AI for Messy Enterprise Data}, 
      author={Fabian Wenz and Felix Treutwein and Kai Arenja and Çagatay Demiralp and Michael Stonebraker},
      year={2026},
      eprint={2604.21413},
      archivePrefix={arXiv},
      primaryClass={cs.DB},
      url={https://arxiv.org/abs/2604.21413}, 
}

@inproceedings{palimpzest,
  author       = {Chunwei Liu and
                  Matthew Russo and
                  Michael J. Cafarella and
                  Lei Cao and
                  Peter Baile Chen and
                  Zui Chen and
                  Michael J. Franklin and
                  Tim Kraska and
                  Samuel Madden and
                  Rana Shahout and
                  Gerardo Vitagliano},
  title        = {Palimpzest: Optimizing AI-Powered Analytics with Declarative Query
                  Processing},
  booktitle    = {15th Conference on Innovative Data Systems Research, {CIDR} 2025},
  year         = {2025}
}

@inproceedings{DBLP:conf/cidr/EckmannB26,
  author       = {Timo Eckmann and
                  Carsten Binnig},
  title        = {A Vision for Autonomous Data Agent Collaboration: From Query-by-Integration
                  to Query-by-Collaboration},
  booktitle    = {16th Conference on Innovative Data Systems Research, {CIDR} 2026},
  year         = {2026}
}

@article{chen2024beaver,
  author       = {Peter Baile Chen and
                  Fabian Wenz and
                  Yi Zhang and
                  Moe Kayali and
                  Nesime Tatbul and
                  Michael J. Cafarella and
                  {\c{C}}agatay Demiralp and
                  Michael Stonebraker},
  title        = {{BEAVER: An Enterprise Benchmark for Text-to-SQL}},
  journal      = {{CoRR}},
  volume       = {abs/2409.02038},
  year         = {2024}
}

@inproceedings{wang2023bird,
  author       = {Jinyang Li and
                  Binyuan Hui and
                  Ge Qu and
                  others},
  title        = {{Can LLM Already Serve as A Database Interface? A Big Bench
                  for Large-Scale Database Grounded Text-to-SQLs}},
  booktitle    = {{Annual Conference on Neural Information Processing 
                  Systems (NeurIPS)}},
  year         = {2023}
}

@inproceedings{yu2018spider,
  author       = {Tao Yu and
                  Rui Zhang and
                  Kai Yang and
                  others},
  title        = {{Spider: A Large-Scale Human-Labeled Dataset for Complex 
                  and Cross-Domain Semantic Parsing and Text-to-SQL Task}},
  booktitle    = {{EMNLP}},
  pages        = {3911--3921},
  year         = {2018}
}
\end{document}